\documentclass[%
 reprint,
 amsmath,amssymb,
 aps,
]{revtex4-1}

\usepackage{graphicx}% Include figure files
\usepackage{dcolumn}% Align table columns on decimal point
\usepackage{bm}% bold math
\usepackage{xcolor}
\begin{document}

\preprint{APS/123-QED}

\title{Geometric optics in the presence of axion-like particles in curved space-time}% Force line breaks with \\

\author{Dominik J. Schwarz}
 \author{Jishnu Goswami}
\author{Aritra Basu}
\affiliation{%
Fakult\"at f\"ur Physik, Universit\"at Bielefeld, Postfach 100131, 33501 Bielefeld, Germany
}

\date{\today}% It is always \today, today,

\begin{abstract}
We present a concise derivation of geometric optics in the 
presence of axionic fields in a curved space-time. Whenever light can 
be described via geometric optics (the eikonal approximation),
the only difference to the situation without axionic field is the 
phenomenon of achromatic birefringence.  Consequently, redshift of light and 
distance estimates based on propagating light rays, as well as shear and magnification due to 
gravitational lensing are not affected by the interaction of light with an axionic field.
\end{abstract}

\maketitle

The study of light propagation in 
the presence of an axion-like field 
in an arbitrary space-time geometry 
is of interest for constrains on
axionic dark matter \cite{IpserSikivie83,SteckerShafi83,Preskill82,Abbott82,Dine82,SikivieYang09} 
and the search for the QCD axion \cite{Kim:2008hd}. 
The latter would solve the strong 
CP problem, but its coupling to 
photons, $\mathsf{g}_{a\gamma}$, remains a free parameter
\cite{sigl18}.

Typically, photon frequencies in astronomical observations and 
optical laboratory experiments are 
much larger than the typical frequency 
of axionic oscillations, which is $\nu_a = 
242\, [m/10^{-21} \,\mathrm{eV}]\, \mathrm{nHz}$
for ultra-light axion-like particles and 
$\nu_a = 242\, [m/\mu\,\mathrm{eV}]\, \mathrm{MHz}$ 
for the QCD axion. Here, $m$ denotes the mass 
of the axionic field. If the corresponding photon wavelength being 
well below all typical length scales of the 
considered application, in particular 
the 
scale of gradients in the axionic field and 
in the structure of space-time, the 
propagation of light can be described in 
the limit of geometric optics (the so-called 
eikonal approximation, \citep[see e.g.,][]{LandauLifshitz}). 

Of late, ultra-light axion-like particles 
are being probed through observations of distant 
astrophysical sources \citep[e.g.,][]{sigl18, fujit19, ivano19, chigusa2020, chen20}. 
Therefore, photon propagation 
in non-Euclidean space-time and their interaction
with axionic fields needs to be established.
In this communication we derive the most 
general equations of geometric optics in 
an arbitrarily curved space-time 
and for an arbitrary configuration of an 
axionic field. 
Our result generalises \cite{HarariSikivie92, Blas_etal19}
to arbitrary space-times and we simplify the derivation 
presented in \cite{Fedderke_etal19}. Another generalisation to 
light coupled to an axionic field and a cold non-magnetized plasma in 
Minkowski space-time is presented in \cite{mcdonald19}.

The signature of the metric tensor
$g_{\mu\nu}$ is chosen to be $(-,+,+,+)$ 
and we use the Heavyside-Lorentz system 
with $\epsilon_0 = \mu_0 = c = \hbar = 1$. 
The action 
\begin{equation}
S = \int d^4 x \,\sqrt{-g}\, {\cal L}
\end{equation}
of the electromagnetic field coupled to an 
axion-like particle (ALP) in a curved 
space-time is given by the Lagrangian (our sign convention for the 
axion-photon coupling follows the one of \cite{HarariSikivie92})
\begin{equation}
{\cal L} = - \frac 14 F_{\mu\nu} F^{\mu\nu} - \frac 12 \partial_\mu a\, \partial^\mu a  + 
\frac{\mathsf{g}_{a\gamma}}{4}\, a\, F_{\mu\nu} \widetilde{F}^{\mu\nu} - V, 
\end{equation}
where $F_{\mu\nu}$ denotes the electromagnetic field strength tensor and $\widetilde{F}^{\mu\nu} = \epsilon^{\mu\nu\alpha\beta} F_{\alpha\beta}/2$ its dual tensor. 
$a$ is an axionic field and $V \equiv V(a)$ its potential. 
Variation with respect to the gauge field $A_\alpha$ gives rise to the equation of motion (using $\nabla_\mu \sqrt{-g} = 0$, where $\nabla_\mu$ denotes the covariant derivative),
\begin{equation}
\nabla_\mu F^{\mu\alpha} - \mathsf{g}_{a\gamma} \,
\epsilon^{\mu\nu\rho\alpha} \,
\partial_\mu A_\nu \, \partial_\rho a = 0 . 
\end{equation}
Making use of $[\nabla_\mu, \nabla_\nu] A^\mu = R_{\mu \nu} A^\mu$, with $R_{\mu \nu}$ denoting the Ricci tensor, and in Lorenz 
gauge, $\nabla_\mu A^\mu = 0$, we obtain,
\begin{equation}
\Box A^\alpha - R^\alpha_\mu\, A^\mu - \mathsf{g}_{a\gamma} \, \epsilon^{\mu\nu\rho\alpha} \, \partial_\mu A_\nu \,
\partial_\rho a = 0. \label{eom}
\end{equation}
Using the rules for exchanging covariant derivatives 
one can see that this equation and the Lorenz gauge are 
invariant under the residual gauge freedom $A^\alpha \to 
A^\alpha + \partial^\alpha f$ with $\Box f = 0$.

The equations of geometric optics 
follow from 
the ansatz $A^\alpha \equiv \mathrm{Re}[{\cal A}^\alpha \exp( \imath \psi)]$ and the assumption that 
large gradients in time and space 
are described by the real phase $\psi$, 
while all other changes are summarised by 
a complex, slowly evolving amplitude ${\cal A}^\alpha$. We define the wave vector  
\begin{equation}
k_\mu \equiv \partial_\mu \psi,
\end{equation}
justified by the resulting equations of motion.
An observer characterised by the space-time 
velocity $u^\mu$ measures the circular
frequency $\omega = - k_\mu u^\mu$.
A direct consequence of the 
definition of the wave vector is
\begin{equation}
\nabla_\mu k_\nu = \nabla_\nu k_\mu, 
\label{eq:hypersurface}
\end{equation}
which reflects the property that the
vector field $k_\mu$ is normal to 
the electromagnetic wave front, which 
defines a hypersurface of space-time.

Inserting the ansatz into the equation of motion in equation~\eqref{eom} and the 
gauge condition, the terms of order $k^2$ and $k$ become,
\begin{eqnarray}
& & k_\mu k^\mu = 0, 
\label{eq:null} \\ 
& & 2\, k^\mu\, \nabla_\mu {\cal A}^\alpha  +  {\cal A}^\alpha \,\nabla_\mu k^\mu -  
\mathsf{g}_{a\gamma} \, \epsilon^{\mu\nu\rho\alpha} \,
k_\mu \, {\cal A}_\nu \,\partial_\rho a = 0, \label{eq:k} \\
&& k_\mu {\cal A}^\mu  = 0, \label{eq:gauge}
\end{eqnarray}  
respectively. The leading 
term in equation~\eqref{eq:null} is the null-condition. Acting on it with a 
covariant derivative and using 
equation~\eqref{eq:hypersurface}, we find the 
equivalent to the geodesic equation
\begin{equation}
k^\mu \nabla_\mu k_\alpha = 0.
\end{equation}
Thus, in the limit of geometric optics and 
in the presence of an axionic field, 
{\em light rays remain to be described by 
null geodesics} in curved space-time. 

We further contract equation~\eqref{eq:k}
with ${\cal A}_\alpha$, define ${\cal A}^2 = {\cal A}_\alpha\,{\cal A}^{\alpha *}$, and obtain,
\begin{equation}
k^\mu \,\nabla_\mu {\cal A}  +  \frac 12 {\cal A}\, \nabla_\mu \, k^\mu = 0, \label{eq:intensity}
\end{equation}
which tells us how the light intensity $I 
= \omega^2 {\cal A}^2$ (see below) changes along the 
light path and shows that {\em the presence of an
axionic field does not affect the observed intensity of a light source}.

Finally, let us introduce the normalised, space-like (complex) polarisation vector $\epsilon^\mu$, i.e.\ $\epsilon^\mu \epsilon_\mu^* = 1$ and 
$\epsilon_\mu k^\mu = 0$, 
and write ${\cal A}^\alpha = {\cal A} \, \epsilon^\alpha$. Returning to equations~\eqref{eq:k} and 
using equation~\eqref{eq:intensity} we find,
\begin{equation}
k^\mu \, \nabla_\mu \, \epsilon^\alpha - \frac 12\,
\mathsf{g}_{a\gamma} \, \epsilon^{\mu\nu\rho\alpha} \, k_\mu\, \epsilon_\nu \,
\partial_\rho a = 0. 
\label{eq:bf1}
\end{equation}
This equation leads to the phenomenon of birefringence. 

To see that most easily, let us introduce 
a orthonormal basis of vectors 
$\{u, e_1, e_2, n\}$, 
consisting of a time-like vector field
$u^\mu$, describing a congruence of 
observers with $u_\mu u^\mu = -1$, a 
space-like vector field $n^\mu \equiv (k^\mu - \omega\, u^\mu)/\omega$, 
pointing from the source to the family of observers,  and 
$n^\mu n_\mu = 1$, and the
space-like linear polarisation basis $e_i^\mu$, $i = 1,2$, 
which spans a screen normal to~$n^\mu$. 
We can use the residual gauge freedom to 
set $\epsilon_\mu n^\mu = \epsilon_\mu u^\mu = 0$.
Then, 
$\epsilon_\mu = \varepsilon_i\, e^i_\mu$, with $e^i_\mu e_j^\mu = \delta^i_j$  
and $|\varepsilon_1|^2 + |\varepsilon_2|^2 = 1$. 
With $k^\mu \,\nabla_\mu e_i^\alpha = 0$ (we parallel 
transport the basis vectors), we can now see
that 
\begin{equation}
(\partial_u + \partial_n) \, \varepsilon^i  - \frac 12\,
\mathsf{g}_{a\gamma}\, [(\partial_u + \partial_n) \, a]\, \epsilon^{u j n i}\, \varepsilon_j  = 0. 
\label{eq:bf2}
\end{equation}
This result is obtained from equation~\eqref{eq:bf1} by expressing 
the wave vector in terms of the base vectors and inserting a 
decomposition of unity in spatial and time-like components
$\delta^\alpha_\beta = h^\alpha_\beta - u^\alpha u_\beta$, 
where $h^\alpha_\beta$ denotes the spatial projection operator,
i.e.~using 
$\delta_\rho^\beta \,\partial_\beta a = h^\beta_\rho \,\partial_\beta a - u_\rho \,\partial_u a$.
The second term in equation~\eqref{eq:bf2} generates a rotation of the polarisation components.
Introducing the coefficients of circular polarisation 
$\varepsilon_{L,R} = (\varepsilon_1 \pm \imath \,\varepsilon_2)/\sqrt{2}$, respectively, we 
obtain two ordinary differential equations
\begin{equation}
\varepsilon_{L,R}^\prime \pm \frac \imath 2\,
\mathsf{g}_{a\gamma}\, a^\prime \,\varepsilon_{L,R}
= 0, 
\end{equation}
where the prime denotes a derivative along the 
line of light propagation. 
The solution to that equation is 
\begin{equation}
\varepsilon_{L,R}(x^\mu_\mathrm{o}) = \exp(\pm \imath \Delta)\, \varepsilon_{L,R}(x^\mu_\mathrm{e}), 
\end{equation}
with $\Delta = \mathsf{g}_{a\gamma}\, [a(x^\mu_\mathrm{e})-a(x^\mu_\mathrm{o})]/2$ is the rotation angle
of the plane of linearly polarized light.
Thus, in the limit of geometric optics, 
{\em birefringence due to axionic fields is achromatic, and depends on the
coupling $\mathsf{g}_{a\gamma}$ and the values of the axionic field at emission and observation.} Moreover,
{\em the curvature of space-time does not affect the birefringence angle $\Delta$, when 
expressed with respect to a parallel transported basis of linear polarisation}.

Chromatic and 
curvature 
effects show up when we go beyond the
eikonal 
approximation, where we would actually return to a wave optics description of the system and no longer talk about light rays. Note that there are no non-linear couplings in this limit. 
It was shown in \cite{Blas_etal19,mcdonald19} that beyond the limit of geometric optics, higher orders in the 
coupling lead to spectral distortions and refractive phenomena show up.

Finally, we can reformulate our results in the 
language of Stokes parameters. 
In the limit of geometric optics and using the same gauge 
fixing as above, the electric field reads 
$E_\alpha \equiv F_{\alpha\beta}\, u^\beta = \mathrm{Re}\,[\omega\, {\cal A}_\alpha\, \exp(\imath\, \psi)]$. 
From that we find the four Stokes parameters as follows: 
\begin{eqnarray}
I &=& \omega^2 {\cal A}^2, \\
Q &=& I \,(|\varepsilon_1|^2 - |\varepsilon_2|^2), \\ 
U &=& 2\, I\, \mathrm{Re}(\varepsilon_1\, \varepsilon_2^*), \\
V &=& - 2\, I\, \mathrm{Im}(\varepsilon_1\, \varepsilon_2^*), 
\end{eqnarray}
{\em This implies that Stokes I remains 
unchanged when photons propagate in axionic field.}

Let us give the example of a 
monochromatic and 
linearly polarised source. For 
such a source, we can write $\epsilon_2(x_e) = r \,
\epsilon_1(x_e)$, with $r$ real, which 
gives
\begin{eqnarray}
Q &=& \frac{I}{1 + r^2}\left[(1-r^2)\, \cos (2\, \Delta) - 2\, r\, \sin(2\, \Delta)\right] ,\\
U &=& \frac{I}{1 + r^2}\left[2\,r\, \cos (2\, \Delta) + (1 - r^2) \sin(2\, \Delta)\right].
\end{eqnarray}
Here, $Q^2 + U^2 = I^2$ and $V=0$.

One could imagine to set up a table top experiment based on finding 
birefringence of a polarised laser beam in vacuum, which could probe 
the mass range of the QCD axion.
In astrophysics the situation of the simple example 
is found in the case of the face-on observation of 
proto-planetary discs \cite{fujit19}. Due to the 
galactic distance scale, ultra-light axions can be 
probed in that case. For other sources with continuous 
emission, the relations for the Stokes parameters may 
be more complicated, but the underlying principle of 
birefringence remains the same. 

We have shown that axionic dark matter
does neither affect the geodesics of light, nor its intensity and therefore
the Etherington theorem \cite{etherington33}, the relation between angular and 
luminosity distance, and the Sachs equations \cite{sachs61}, 
which describe the propagation of light bundles in 
curved space-times, also hold true in the presence of axion-photon 
interaction and in the limit of geometric optics. Thus, 
{\em cosmic distance estimation, based on parallaxes, standard candles or standard rulers are not affected. Also the photon-axion interaction cannot change the redshift of a source.} Our result further 
implies, that the presence of axionic dark matter does 
not modify the conclusions on gravitational 
lensing as long as shape and magnification are concerned.

On the other hand, axionic dark matter gives rise to an 
additional achromatic birefringence that adds to the 
chromatic birefringence introduced by the cosmic 
magneto-ionic medium \cite{Carilli02, Akahori10}.
The investigation of the interplay of a magnetized plasma with light propagation in an axionic field in flat space-time \cite{mcdonald19} shows that we should 
also expect additional effects of the order of the plasma frequency squared times the axion-photon coupling, which are subdominant. Thus, the 
achromatic nature of axionic birefringence might provide a smoking gun in the search for ultra-light axionic 
dark matter on cosmological and astrophysical length and time scales.

\begin{acknowledgments}
We thank Yuko Urakawa for very helpful discussions and comments and Walter Pfeiffer for discussions regarding future implications on laboratory 
experiments. We acknowledge support by the Deutsche Forschungsgemeinschaft 
(DFG, German Research Foundation) through the CRC-TR 211 `Strong-interaction matter under extreme conditions'– project number 315477589 – TRR 211 and by the German Federal Ministry
of Education and Research (BMBF) under grant 05A17PB1 (Verbundprojekt
D-MeerKAT).
\end{acknowledgments}

\bibliography{ALP_geometric_optics}

\begin{thebibliography}{21}%
\makeatletter
\providecommand \@ifxundefined [1]{%
 \@ifx{#1\undefined}
}%
\providecommand \@ifnum [1]{%
 \ifnum #1\expandafter \@firstoftwo
 \else \expandafter \@secondoftwo
 \fi
}%
\providecommand \@ifx [1]{%
 \ifx #1\expandafter \@firstoftwo
 \else \expandafter \@secondoftwo
 \fi
}%
\providecommand \natexlab [1]{#1}%
\providecommand \enquote  [1]{``#1''}%
\providecommand \bibnamefont  [1]{#1}%
\providecommand \bibfnamefont [1]{#1}%
\providecommand \citenamefont [1]{#1}%
\providecommand \href@noop [0]{\@secondoftwo}%
\providecommand \href [0]{\begingroup \@sanitize@url \@href}%
\providecommand \@href[1]{\@@startlink{#1}\@@href}%
\providecommand \@@href[1]{\endgroup#1\@@endlink}%
\providecommand \@sanitize@url [0]{\catcode `\\12\catcode `\$12\catcode
  `\&12\catcode `\#12\catcode `\^12\catcode `\_12\catcode `\%12\relax}%
\providecommand \@@startlink[1]{}%
\providecommand \@@endlink[0]{}%
\providecommand \url  [0]{\begingroup\@sanitize@url \@url }%
\providecommand \@url [1]{\endgroup\@href {#1}{\urlprefix }}%
\providecommand \urlprefix  [0]{URL }%
\providecommand \Eprint [0]{\href }%
\providecommand \doibase [0]{http://dx.doi.org/}%
\providecommand \selectlanguage [0]{\@gobble}%
\providecommand \bibinfo  [0]{\@secondoftwo}%
\providecommand \bibfield  [0]{\@secondoftwo}%
\providecommand \translation [1]{[#1]}%
\providecommand \BibitemOpen [0]{}%
\providecommand \bibitemStop [0]{}%
\providecommand \bibitemNoStop [0]{.\EOS\space}%
\providecommand \EOS [0]{\spacefactor3000\relax}%
\providecommand \BibitemShut  [1]{\csname bibitem#1\endcsname}%
\let\auto@bib@innerbib\@empty
%</preamble>
\bibitem [{\citenamefont {Ipser}\ and\ \citenamefont
  {Sikivie}(1983)}]{IpserSikivie83}%
  \BibitemOpen
  \bibfield  {author} {\bibinfo {author} {\bibfnamefont {J.}~\bibnamefont
  {Ipser}}\ and\ \bibinfo {author} {\bibfnamefont {P.}~\bibnamefont
  {Sikivie}},\ }\href {\doibase 10.1103/PhysRevLett.50.925} {\bibfield
  {journal} {\bibinfo  {journal} {Phys. Rev. Lett.}\ }\textbf {\bibinfo
  {volume} {50}},\ \bibinfo {pages} {925} (\bibinfo {year} {1983})}\BibitemShut
  {NoStop}%
%%CITATION = PRLTA,50,925;%%
\bibitem [{\citenamefont {Stecker}\ and\ \citenamefont
  {Shafi}(1983)}]{SteckerShafi83}%
  \BibitemOpen
  \bibfield  {author} {\bibinfo {author} {\bibfnamefont {F.~W.}\ \bibnamefont
  {Stecker}}\ and\ \bibinfo {author} {\bibfnamefont {Q.}~\bibnamefont
  {Shafi}},\ }\href {\doibase 10.1103/PhysRevLett.50.928} {\bibfield  {journal}
  {\bibinfo  {journal} {Phys. Rev. Lett.}\ }\textbf {\bibinfo {volume} {50}},\
  \bibinfo {pages} {928} (\bibinfo {year} {1983})}\BibitemShut {NoStop}%
%%CITATION = PRLTA,50,928;%%
\bibitem [{\citenamefont {Preskill}\ \emph {et~al.}(1983)\citenamefont
  {Preskill}, \citenamefont {Wise},\ and\ \citenamefont
  {Wilczek}}]{Preskill82}%
  \BibitemOpen
  \bibfield  {author} {\bibinfo {author} {\bibfnamefont {J.}~\bibnamefont
  {Preskill}}, \bibinfo {author} {\bibfnamefont {M.~B.}\ \bibnamefont {Wise}},
  \ and\ \bibinfo {author} {\bibfnamefont {F.}~\bibnamefont {Wilczek}},\ }\href
  {\doibase 10.1016/0370-2693(83)90637-8} {\bibfield  {journal} {\bibinfo
  {journal} {Phys. Lett.}\ }\textbf {\bibinfo {volume} {120B}},\ \bibinfo
  {pages} {127} (\bibinfo {year} {1983})}\BibitemShut {NoStop}%
%%CITATION = PHLTA,120B,127;%%
\bibitem [{\citenamefont {Abbott}\ and\ \citenamefont
  {Sikivie}(1983)}]{Abbott82}%
  \BibitemOpen
  \bibfield  {author} {\bibinfo {author} {\bibfnamefont {L.~F.}\ \bibnamefont
  {Abbott}}\ and\ \bibinfo {author} {\bibfnamefont {P.}~\bibnamefont
  {Sikivie}},\ }\href {\doibase 10.1016/0370-2693(83)90638-X} {\bibfield
  {journal} {\bibinfo  {journal} {Phys. Lett.}\ }\textbf {\bibinfo {volume}
  {120B}},\ \bibinfo {pages} {133} (\bibinfo {year} {1983})}\BibitemShut
  {NoStop}%
%%CITATION = PHLTA,120B,133;%%
\bibitem [{\citenamefont {Dine}\ and\ \citenamefont {Fischler}(1983)}]{Dine82}%
  \BibitemOpen
  \bibfield  {author} {\bibinfo {author} {\bibfnamefont {M.}~\bibnamefont
  {Dine}}\ and\ \bibinfo {author} {\bibfnamefont {W.}~\bibnamefont
  {Fischler}},\ }\href {\doibase 10.1016/0370-2693(83)90639-1} {\bibfield
  {journal} {\bibinfo  {journal} {Phys. Lett.}\ }\textbf {\bibinfo {volume}
  {120B}},\ \bibinfo {pages} {137} (\bibinfo {year} {1983})}\BibitemShut
  {NoStop}%
%%CITATION = PHLTA,120B,137;%%
\bibitem [{\citenamefont {Sikivie}\ and\ \citenamefont
  {Yang}(2009)}]{SikivieYang09}%
  \BibitemOpen
  \bibfield  {author} {\bibinfo {author} {\bibfnamefont {P.}~\bibnamefont
  {Sikivie}}\ and\ \bibinfo {author} {\bibfnamefont {Q.}~\bibnamefont {Yang}},\
  }\href {\doibase 10.1103/PhysRevLett.103.111301} {\bibfield  {journal}
  {\bibinfo  {journal} {Phys. Rev. Lett.}\ }\textbf {\bibinfo {volume} {103}},\
  \bibinfo {pages} {111301} (\bibinfo {year} {2009})},\ \Eprint
  {http://arxiv.org/abs/0901.1106} {arXiv:0901.1106 [hep-ph]} \BibitemShut
  {NoStop}%
%%CITATION = ARXIV:0901.1106;%%
\bibitem [{\citenamefont {Kim}\ and\ \citenamefont
  {Carosi}(2010)}]{Kim:2008hd}%
  \BibitemOpen
  \bibfield  {author} {\bibinfo {author} {\bibfnamefont {J.~E.}\ \bibnamefont
  {Kim}}\ and\ \bibinfo {author} {\bibfnamefont {G.}~\bibnamefont {Carosi}},\
  }\href {\doibase 10.1103/RevModPhys.82.557,10.1103/RevModPhys.91.049902,
  10.1103/RevModPhys.91.049902, 10.1103/RevModPhys.82.557} {\bibfield
  {journal} {\bibinfo  {journal} {Rev. Mod. Phys.}\ }\textbf {\bibinfo {volume}
  {82}},\ \bibinfo {pages} {557} (\bibinfo {year} {2010})},\ \bibinfo {note}
  {[erratum: Rev. Mod. Phys.91,no.4,049902(2019)]},\ \Eprint
  {http://arxiv.org/abs/0807.3125} {arXiv:0807.3125 [hep-ph]} \BibitemShut
  {NoStop}%
%%CITATION = ARXIV:0807.3125;%%
\bibitem [{\citenamefont {Sigl}\ and\ \citenamefont {Trivedi}(2018)}]{sigl18}%
  \BibitemOpen
  \bibfield  {author} {\bibinfo {author} {\bibfnamefont {G.}~\bibnamefont
  {Sigl}}\ and\ \bibinfo {author} {\bibfnamefont {P.}~\bibnamefont {Trivedi}},\
  }\href@noop {} {\  (\bibinfo {year} {2018})},\ \Eprint
  {http://arxiv.org/abs/1811.07873} {arXiv:1811.07873 [astro-ph.CO]}
  \BibitemShut {NoStop}%
%%CITATION = ARXIV:1811.07873;%%
\bibitem [{\citenamefont {Landau}\ and\ \citenamefont
  {Lifshitz}(1987)}]{LandauLifshitz}%
  \BibitemOpen
  \bibfield  {author} {\bibinfo {author} {\bibfnamefont {L.~D.}\ \bibnamefont
  {Landau}}\ and\ \bibinfo {author} {\bibfnamefont {E.~M.}\ \bibnamefont
  {Lifshitz}},\ }\href@noop {} {\emph {\bibinfo {title} {Klassische
  Feldtheorie, Lehrbuch der Theoretischen Physik, Band 2}}}\ (\bibinfo
  {publisher} {Akademie-Verlag},\ \bibinfo {address} {Berlin},\ \bibinfo {year}
  {1987})\BibitemShut {NoStop}%
\bibitem [{\citenamefont {Fujita}\ \emph {et~al.}(2019)\citenamefont {Fujita},
  \citenamefont {Tazaki},\ and\ \citenamefont {Toma}}]{fujit19}%
  \BibitemOpen
  \bibfield  {author} {\bibinfo {author} {\bibfnamefont {T.}~\bibnamefont
  {Fujita}}, \bibinfo {author} {\bibfnamefont {R.}~\bibnamefont {Tazaki}}, \
  and\ \bibinfo {author} {\bibfnamefont {K.}~\bibnamefont {Toma}},\ }\href
  {\doibase 10.1103/PhysRevLett.122.191101} {\bibfield  {journal} {\bibinfo
  {journal} {Phys. Rev. Lett.}\ }\textbf {\bibinfo {volume} {122}},\ \bibinfo
  {pages} {191101} (\bibinfo {year} {2019})},\ \Eprint
  {http://arxiv.org/abs/1811.03525} {arXiv:1811.03525 [astro-ph.CO]}
  \BibitemShut {NoStop}%
%%CITATION = ARXIV:1811.03525;%%
\bibitem [{\citenamefont {{Ivanov}}\ \emph {et~al.}(2019)\citenamefont
  {{Ivanov}}, \citenamefont {{Kovalev}}, \citenamefont {{Lister}},
  \citenamefont {{Panin}}, \citenamefont {{Pushkarev}}, \citenamefont
  {{Savolainen}},\ and\ \citenamefont {{Troitsky}}}]{ivano19}%
  \BibitemOpen
  \bibfield  {author} {\bibinfo {author} {\bibfnamefont {M.~M.}\ \bibnamefont
  {{Ivanov}}}, \bibinfo {author} {\bibfnamefont {Y.~Y.}\ \bibnamefont
  {{Kovalev}}}, \bibinfo {author} {\bibfnamefont {M.~L.}\ \bibnamefont
  {{Lister}}}, \bibinfo {author} {\bibfnamefont {A.~G.}\ \bibnamefont
  {{Panin}}}, \bibinfo {author} {\bibfnamefont {A.~B.}\ \bibnamefont
  {{Pushkarev}}}, \bibinfo {author} {\bibfnamefont {T.}~\bibnamefont
  {{Savolainen}}}, \ and\ \bibinfo {author} {\bibfnamefont {S.~V.}\
  \bibnamefont {{Troitsky}}},\ }\href {\doibase 10.1088/1475-7516/2019/02/059}
  {\bibfield  {journal} {\bibinfo  {journal} {JCAP}\ }\textbf {\bibinfo
  {volume} {2019}},\ \bibinfo {eid} {059} (\bibinfo {year} {2019})},\ \Eprint
  {http://arxiv.org/abs/1811.10997} {arXiv:1811.10997 [astro-ph.CO]}
  \BibitemShut {NoStop}%
\bibitem [{\citenamefont {Chigusa}\ \emph {et~al.}(2020)\citenamefont
  {Chigusa}, \citenamefont {Moroi},\ and\ \citenamefont
  {Nakayama}}]{chigusa2020}%
  \BibitemOpen
  \bibfield  {author} {\bibinfo {author} {\bibfnamefont {S.}~\bibnamefont
  {Chigusa}}, \bibinfo {author} {\bibfnamefont {T.}~\bibnamefont {Moroi}}, \
  and\ \bibinfo {author} {\bibfnamefont {K.}~\bibnamefont {Nakayama}},\ }\href
  {\doibase 10.1016/j.physletb.2020.135288} {\bibfield  {journal} {\bibinfo
  {journal} {Phys. Lett.}\ }\textbf {\bibinfo {volume} {B803}},\ \bibinfo
  {pages} {135288} (\bibinfo {year} {2020})},\ \Eprint
  {http://arxiv.org/abs/1911.09850} {arXiv:1911.09850 [astro-ph.CO]}
  \BibitemShut {NoStop}%
%%CITATION = ARXIV:1911.09850;%%
\bibitem [{\citenamefont {Chen}\ \emph {et~al.}(2020)\citenamefont {Chen},
  \citenamefont {Shu}, \citenamefont {Xue}, \citenamefont {Yuan},\ and\
  \citenamefont {Zhao}}]{chen20}%
  \BibitemOpen
  \bibfield  {author} {\bibinfo {author} {\bibfnamefont {Y.}~\bibnamefont
  {Chen}}, \bibinfo {author} {\bibfnamefont {J.}~\bibnamefont {Shu}}, \bibinfo
  {author} {\bibfnamefont {X.}~\bibnamefont {Xue}}, \bibinfo {author}
  {\bibfnamefont {Q.}~\bibnamefont {Yuan}}, \ and\ \bibinfo {author}
  {\bibfnamefont {Y.}~\bibnamefont {Zhao}},\ }\href {\doibase
  10.1103/PhysRevLett.124.061102} {\bibfield  {journal} {\bibinfo  {journal}
  {Phys. Rev. Lett.}\ }\textbf {\bibinfo {volume} {124}},\ \bibinfo {pages}
  {061102} (\bibinfo {year} {2020})},\ \Eprint
  {http://arxiv.org/abs/1905.02213} {arXiv:1905.02213 [hep-ph]} \BibitemShut
  {NoStop}%
%%CITATION = ARXIV:1905.02213;%%
\bibitem [{\citenamefont {Harari}\ and\ \citenamefont
  {Sikivie}(1992)}]{HarariSikivie92}%
  \BibitemOpen
  \bibfield  {author} {\bibinfo {author} {\bibfnamefont {D.}~\bibnamefont
  {Harari}}\ and\ \bibinfo {author} {\bibfnamefont {P.}~\bibnamefont
  {Sikivie}},\ }\href {\doibase 10.1016/0370-2693(92)91363-E} {\bibfield
  {journal} {\bibinfo  {journal} {Phys. Lett.}\ }\textbf {\bibinfo {volume}
  {B289}},\ \bibinfo {pages} {67} (\bibinfo {year} {1992})}\BibitemShut
  {NoStop}%
%%CITATION = PHLTA,B289,67;%%
\bibitem [{\citenamefont {Blas}\ \emph {et~al.}(2020)\citenamefont {Blas},
  \citenamefont {Caputo}, \citenamefont {Ivanov},\ and\ \citenamefont
  {Sberna}}]{Blas_etal19}%
  \BibitemOpen
  \bibfield  {author} {\bibinfo {author} {\bibfnamefont {D.}~\bibnamefont
  {Blas}}, \bibinfo {author} {\bibfnamefont {A.}~\bibnamefont {Caputo}},
  \bibinfo {author} {\bibfnamefont {M.~M.}\ \bibnamefont {Ivanov}}, \ and\
  \bibinfo {author} {\bibfnamefont {L.}~\bibnamefont {Sberna}},\ }\href
  {\doibase 10.1016/j.dark.2019.100428} {\bibfield  {journal} {\bibinfo
  {journal} {Physics of the Dark Universe}\ }\textbf {\bibinfo {volume} {27}},\
  \bibinfo {pages} {100428} (\bibinfo {year} {2020})}\BibitemShut {NoStop}%
\bibitem [{\citenamefont {Fedderke}\ \emph {et~al.}(2019)\citenamefont
  {Fedderke}, \citenamefont {Graham},\ and\ \citenamefont
  {Rajendran}}]{Fedderke_etal19}%
  \BibitemOpen
  \bibfield  {author} {\bibinfo {author} {\bibfnamefont {M.~A.}\ \bibnamefont
  {Fedderke}}, \bibinfo {author} {\bibfnamefont {P.~W.}\ \bibnamefont
  {Graham}}, \ and\ \bibinfo {author} {\bibfnamefont {S.}~\bibnamefont
  {Rajendran}},\ }\href {\doibase 10.1103/PhysRevD.100.015040} {\bibfield
  {journal} {\bibinfo  {journal} {Phys. Rev.}\ }\textbf {\bibinfo {volume}
  {D100}},\ \bibinfo {pages} {015040} (\bibinfo {year} {2019})},\ \Eprint
  {http://arxiv.org/abs/1903.02666} {arXiv:1903.02666 [astro-ph.CO]}
  \BibitemShut {NoStop}%
%%CITATION = ARXIV:1903.02666;%%
\bibitem [{\citenamefont {McDonald}\ and\ \citenamefont
  {Ventura}(2019)}]{mcdonald19}%
  \BibitemOpen
  \bibfield  {author} {\bibinfo {author} {\bibfnamefont {J.~I.}\ \bibnamefont
  {McDonald}}\ and\ \bibinfo {author} {\bibfnamefont {L.~B.}\ \bibnamefont
  {Ventura}},\ }\href@noop {} {\  (\bibinfo {year} {2019})},\ \Eprint
  {http://arxiv.org/abs/1911.10221} {arXiv:1911.10221 [hep-ph]} \BibitemShut
  {NoStop}%
%%CITATION = ARXIV:1911.10221;%%
\bibitem [{\citenamefont {{Etherington}}(1933)}]{etherington33}%
  \BibitemOpen
  \bibfield  {author} {\bibinfo {author} {\bibfnamefont {I.~M.~H.}\
  \bibnamefont {{Etherington}}},\ }\href@noop {} {\bibfield  {journal}
  {\bibinfo  {journal} {Philosophical Magazine}\ }\textbf {\bibinfo {volume}
  {15}},\ \bibinfo {pages} {761} (\bibinfo {year} {1933})}\BibitemShut
  {NoStop}%
\bibitem [{\citenamefont {{Sachs}}(1961)}]{sachs61}%
  \BibitemOpen
  \bibfield  {author} {\bibinfo {author} {\bibfnamefont {R.}~\bibnamefont
  {{Sachs}}},\ }\href {\doibase 10.1098/rspa.1961.0202} {\bibfield  {journal}
  {\bibinfo  {journal} {Proc. Roy. Soc. Lond. Ser. A}\ }\textbf {\bibinfo
  {volume} {264}},\ \bibinfo {pages} {309} (\bibinfo {year}
  {1961})}\BibitemShut {NoStop}%
\bibitem [{\citenamefont {{Carilli}}\ and\ \citenamefont
  {{Taylor}}(2002)}]{Carilli02}%
  \BibitemOpen
  \bibfield  {author} {\bibinfo {author} {\bibfnamefont {C.~L.}\ \bibnamefont
  {{Carilli}}}\ and\ \bibinfo {author} {\bibfnamefont {G.~B.}\ \bibnamefont
  {{Taylor}}},\ }\href {\doibase 10.1146/annurev.astro.40.060401.093852}
  {\bibfield  {journal} {\bibinfo  {journal} {ARA\&A}\ }\textbf {\bibinfo
  {volume} {40}},\ \bibinfo {pages} {319} (\bibinfo {year} {2002})},\ \Eprint
  {http://arxiv.org/abs/astro-ph/0110655} {arXiv:astro-ph/0110655 [astro-ph]}
  \BibitemShut {NoStop}%
\bibitem [{\citenamefont {{Akahori}}\ and\ \citenamefont
  {{Ryu}}(2010)}]{Akahori10}%
  \BibitemOpen
  \bibfield  {author} {\bibinfo {author} {\bibfnamefont {T.}~\bibnamefont
  {{Akahori}}}\ and\ \bibinfo {author} {\bibfnamefont {D.}~\bibnamefont
  {{Ryu}}},\ }\href {\doibase 10.1088/0004-637X/723/1/476} {\bibfield
  {journal} {\bibinfo  {journal} {ApJ}\ }\textbf {\bibinfo {volume} {723}},\
  \bibinfo {pages} {476} (\bibinfo {year} {2010})},\ \Eprint
  {http://arxiv.org/abs/1009.0570} {arXiv:1009.0570 [astro-ph.CO]} \BibitemShut
  {NoStop}%
\end{thebibliography}%

\end{document}